\documentclass[onecolumn, 12pt]{IEEEtran_stretto3}
\usepackage[dvips]{graphicx}
\def\BibTeX{{\rm B\kern-.05em{\sc i\kern-.02
5em b}\kern-.08em\kern-.1667em\lower.7ex\hbox{E}\kern-.125emX}}

\begin{document}
\title{Design of optimal convolutional codes for joint decoding of correlated sources in wireless sensor networks}
\author{Andrea Abrardo}
\maketitle

\begin{center}
\emph{Department of Information Engineering - University of Siena
\\ Via Roma, 56 - 53100 Siena, ITALY}\\

\vspace{0.3 cm} $\mathbf{Contact ~Author:}$ Andrea Abrardo
\\ e-mail: abrardo@ing.unisi.it, Tel.: +39 0577 234624, Fax: +39
0577 233602
\end{center}

\markboth{A. Abrardo, "Design of optimal convolutional codes for
joint decoding of correlated sources in wireless sensor networks"}{}

\begin{abstract}
We consider a wireless sensors network scenario where two nodes
detect correlated sources and deliver them to a central collector
via a wireless link. Differently from the Slepian-Wolf approach to
distributed source coding, in the proposed scenario the sensing
nodes do not perform any pre-compression of the sensed data.
Original data are instead independently encoded by means of
low-complexity convolutional codes. The decoder performs joint
decoding with the aim of exploiting the inherent correlation between
the transmitted sources. Complexity at the decoder is kept low
thanks to the use of an iterative joint decoding scheme, where the
output of each decoder is fed to the other decoder's input as
a-priori information. For such scheme, we derive a novel analytical
framework for evaluating an upper bound of joint-detection packet
error probability and for deriving the optimum coding scheme.
Experimental results confirm the validity of the analytical
framework, and show that recursive codes allow a noticeable
performance gain with respect to non-recursive coding schemes.
Moreover, the proposed recursive coding scheme allows to approach
the ideal Slepian-Wolf scheme performance in AWGN channel, and to
clearly outperform it
over fading channels on account of diversity gain due to correlation of information.

\end{abstract}

\section{Introduction}

Wireless sensor networks have recently received a lot of attention
in the research literature \cite{Akyildiz}. The efficient
transmission of correlated signals observed at different nodes to
one or more collectors, is one of the main challenges in such
networks. In the case of one collector node, this problem is often
referred to as reach-back channel in the literature \cite{Barros},
\cite{Gupta}, \cite{Gamal}. In its most simple form, the problem can
be summarized as follows: two independent nodes have to transmit
correlated sensed data to a collector node by using the minimum
energy, i.e., by exploiting in some way the implicit correlation
among data. In an attempt to exploit such correlation, many works
have recently focussed on the design of coding schemes that approach
the Slepian-Wolf fundamental limit on the achievable compression
rates \cite{Aaron}, \cite{Bajcsy}, \cite{Deslauriers}, \cite{Xiong}.
However, approaching the Slepian-Wolf compression limit requires in
general a huge implementation complexity at the transmitter (in
terms of number of operations and memory requirements) that in many
cases is not compatible with the needs of deploying very
light-weight, low cost, and low consuming sensor nodes. Alternative
approaches to distributed source coding are represented by
cooperative source-channel coding schemes and joint source-channel
coding.

In a cooperative system, each user is assigned one or more partners.
The partners overhear each other's transmitted signals, process
these signals, and retransmit toward the destination to provide
extra observations of the source signal at the collector. Even
though the inter partner channel is noisy, the virtual
transmit-antenna array consisting of these partners provides
additional diversity, and may entail improvements in terms of error
rates and throughput for all the nodes involved \cite{Stefanov1},
\cite{Stefanov2}, \cite{Sendonaris1}, \cite{Sendonaris2}
\cite{Laneman1}, \cite{Laneman2}. This approach can take advantage
of correlation among the different information flows simply by
including Slepian-Wolf based source coding schemes, i.e., the
sensing nodes transmit compressed version of the sensed data each
other, so that cooperative source-channel coding schemes can be
derived \cite{Gamal1}. However, approaches based on cooperation
require a strict coordination/synchronization among nodes, so that
they can be considered as a single transmitter equipped with
multiple antennas. This entails a more complex design of low level
protocols and forces the nodes to fully decode signals from the
other nodes. This operation is of course power consuming, and in
some cases such an additional power can partially or completely
eliminate the advantage of distributed diversity.

An alternative solution to exploit correlation among users is
represented by joint source-channel coding. In this case, no
cooperation among nodes is required and the correlated sources are
not source encoded but only channel encoded at a reduced rate (with
respect to the uncorrelated case). The reduced reliability due to
channel coding rate reduction can be compensated by exploiting
intrinsic correlation among different information sources at the
channel decoder. Such an approach has attracted the attention of
several researchers in the recent past on account of its
implementation simplicity \cite{Garcia-Frias}, \cite{Mondin1},
\cite{Maramatsu}, \cite{Mondin2}. Works dealing with joint
source-channel coding have so far considered classical turbo or LDPC
codes, where the decoder can exploit the correlation among sources
by performing message passing between the two decoders. However, in
order to exploit the potentialities of such codes it is necessary to
envisage very long transmitted sequences (often in the order of
10000 bits or even longer), a situation which is not so common in
wireless sensor networks' applications where in general the nodes
have to deliver a small packet of bits. Of course, the same encoding
and decoding principles of turbo/LDPC codes can be used with shorter
block lengths, but the decoder's performance becomes in this case
similar to that of classical block or convolutional codes.



In this paper, we will consider a joint source-channel coding scheme
based on a low-complexity (i.e., small number of states)
convolutional coding scheme. In this case, both the memory
requirement at the encoder and the transmission delay are of very
few bits (i.e., the constraint length of the code). Moreover,
similarly to turbo or LDPC schemes, the complexity at the decoder
can be kept low thanks to the use of an iterative joint decoding
scheme, where the output of each decoder is fed to the other
decoder's input as a-priori information. It is worth noting that
when a convolutional code is used to provide forward error
correction for packet data transmissions, we are in general
interested in the average probability of block (or packet)
error rather than in the bit error rate \cite{Lassing}. \\
In order to manage the problem complexity, we assume that a-priori
information is ideal, i.e., it is identical to the original
information transmitted by the other encoder. In this case, the
correlation between the a-priori information and the to-be-decoded
bits is still equal to the original correlation between the
information signals, and the problem turns out to be that of Viterbi
decoding with a-priori soft information.\\
To the best of my knowledge, the first paper which studies this
problem is an old paper by Hagenauer \cite{Hagenauer}. The bounds
found by Hagenauer are generally accepted by the research community,
and a recent paper \cite{Mondinf} uses such bounds to evaluate the
performance of a joint convolutional decoding system similar to the
one proposed in this paper. Unfortunately, the bounds found by
Hagenauer are far from being satisfying, as we will show in Section
IV. In particular, in \cite{Hagenauer} it is assumed a perfect match
between the a-priori information hard decision parameter, i.e., the
sign of the a-priori log-likelihood values, and the actually
transmitted information signal. On the other hand, in \cite{Mondinf}
the good match between simulations and theoretical curves is due to
the use of base-10 logarithm instead of the correct natural
logarithm. Hence, this paper removes the assumptions made in
\cite{Hagenauer} and a novel analytical framework, where the packet
error probability is evaluated by averaging over all possible
configuration of a-priori information, is provided. Such an analysis
is then considered for deriving optimal coding schemes for the
scenario proposed in this paper.

This paper is organized as follows. Section II describes the
proposed scenario and gives notations used throughout the rest of
the paper. In Section III, starting from the definition of the
optimum MAP joint-decoding problem, we derive a sub-optimum
iterative joint-decoding scheme. Section IV and V illustrate the
analysis which allows to evaluate the packet error probabilities of
convolutional joint-decoding and to derive the optimum code
searching strategy. Finally, Section VI shows results and
comparisons.

\section{Scenario}

\begin{figure}
\begin{center}
\includegraphics[width=0.5\textwidth]{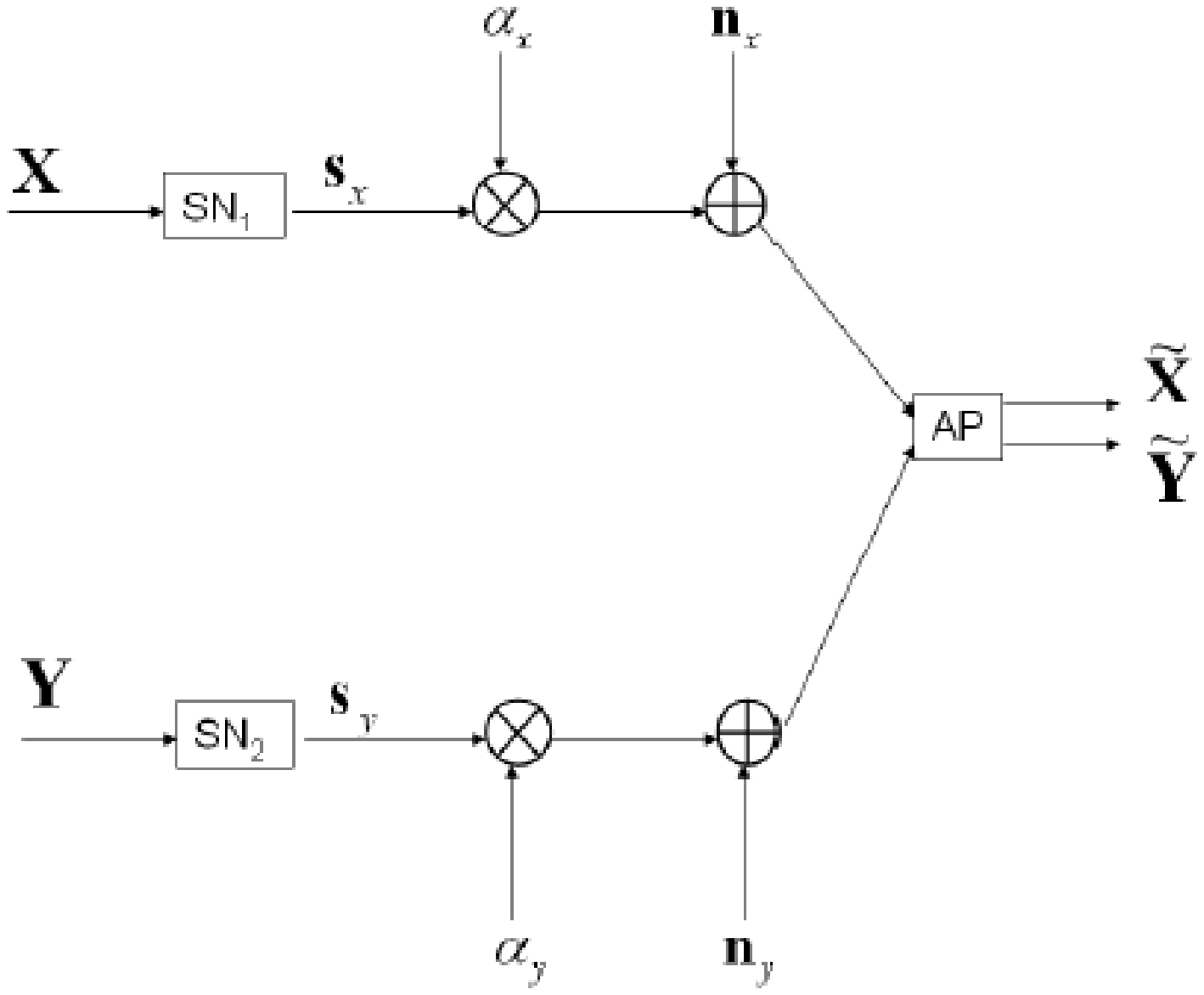}
\end{center} \caption{The proposed two sensing nodes scenario} \label{Fig1}
\end{figure}

Let's consider the detecting problem shown in Figure \ref{Fig1}. We
have two sensor nodes, namely $SN_{1}$ and $SN_{2}$, which detect
the two binary correlated signals \textbf{{X}} and \textbf{Y},
respectively. Such signals, referred to as information signals in
the following, are taken to be i.i.d. correlated binary randon
variables with $P_r\left\{x_i = 1/0\right\} = P_r\left\{y_i =
1/0\right\} = 0.5$ and correlation $\rho = P_r\left\{x_i =
y_i\right\} > 0.5$.\\
The information signals, which are assumed to be detectable without
error (i.e., ideal sensor nodes), must be delivered to the access
point node (AP). To this aim, sensor nodes can establish a direct
link toward the AP. We assume that the communication links are
affected by independent link gains and by additive AWGN noise.
Referring to the vectorial equivalent low-pass signal
representation, we denote to as ${\mathbf{s}}$ the complex
transmitted vector which conveys the information signal, $\alpha$
the complex link gain term which encompasses both path loss and
fading, and ${\mathbf{n}}$ the complex additive noise. As for the
channel model, we assume an almost static system characterized by
very slow fading, so that the channel link gains can be perfectly
estimated at the receiver \footnote{This assumption is reasonable
since in most wireless sensor networks' applications sensor nodes
are static or almost
static}.\\
Let's assume that each transmitter uses a rate $r = k/n$ binary
antipodal channel coding scheme to protect information from channel
errors, and denote to as ${\mathbf{x}} =
\left(x_0,x_1,\ldots,x_{k-1}\right)$ and ${\mathbf{z}} =
\left(z_0,z_1,\ldots,z_{n-1}\right)$, with $z_i = \pm 1$, the
information and the coded sequences for $SN_{1}$, respectively. In
an analogous manner, ${\mathbf{y}} =
\left(y_0,y_1,\ldots,y_{k-1}\right)$ and ${\mathbf{w}} =
\left(w_0,w_1,\ldots,w_{n-1}\right)$, with $w_i = \pm 1$, are the
information and the coded sequences for
$SN_{2}$.\\
Eventually, let's denote to as $E(\cdot)$ the mean operator and
introduce the following terms: $\xi_{x} =
E\left(\left|{\mathbf{s}}_x\right|^2/2\right)$, is the energy per
coded sample transmitted by $SN_1$, $\xi_{y} =
E\left(\left|{\mathbf{s}}_y\right|^2/2\right)$, is the energy per
coded sample transmitted by $SN_2$, $G_x =
\left|\alpha_{x}\right|^2$, is the power gain term for the first
link, $G_y = \left|\alpha_{y}\right|^2$, is the power gain term for
the second link, $E\left(\left|{\mathbf{n}}_x\right|^2\right) =
E\left(\left|{\mathbf{n}}_y\right|^2\right) = 2N_0$, is the variance
of the AWGN noise.

The coded sequence is transmitted into the channel with an antipodal
binary modulation scheme (PSK), i.e., $s_{x,i} = z_i
\sqrt{2\xi_{x}}$, $s_{y,i} = w_i \sqrt{2\xi_{y}}$. Hence, denoting
to as $u_{x,i}$ and $u_{y,i}$ the decision variable at the receiver,
we get:

\begin{equation}
\begin{array}{c}
u_{i,x} = z_i \sqrt{2 G_x \xi_{x}}   + \eta_{i,x}\\
u_{i,y} = w_i \sqrt{2 G_y \xi_{y}}   + \eta_{i,y}
\end{array}
 \label{eq1AA}
\end{equation}
where $\eta_{i,x}$, $\eta_{i,y}$ are Gaussian random noise terms
with zero mean and variance $N_0$. The energy per information bit
for the two links can be written as $\xi_{b,x} = \frac{G_x
\xi_{x}}{r}$ and $\xi_{b,y} = \frac{G_y \xi_{y}}{r}$, respectively.
Denoting to as $\xi_{c,x}=r \xi_{b,x}$ and $\xi_{c,y}=r \xi_{b,y}$
the received energy per coded bit for the two links, we can rewrite
equation (\ref{eq1AA}) as:
\begin{equation}
\begin{array}{c}
u_{i,x} = z_i \sqrt{2 \xi_{c,x}}   + \eta_{i,x}\\
u_{i,y} = w_i \sqrt{2 \xi_{c,y}}   + \eta_{i,y}
\end{array}
 \label{eq2AA}
\end{equation}
 Note that the same model attains also for a more
efficient quaternary modulation scheme (QPSK), where two coded
symbols are transmitted at the same time in the real and imaginary
part of the complex transmitted sample.

\section{Iterative joint-decoding}

The decoders' problem is that of providing an estimation of
${\mathbf{x}}$ and ${\mathbf{y}}$ given the observation sequences
${\mathbf{u}}_x$ and ${\mathbf{u}}_y$. Since ${\mathbf{x}}$ and
${\mathbf{y}}$ are correlated, the optimum decoding problem can be
addressed as a MAP joint decoding problem:
\begin{equation}
    \begin{array}{cl}
    \left\{\tilde{{\mathbf{x}}},\tilde{{\mathbf{y}}}\right\} = \mathop{arg ~
max}\limits_{{{\mathbf{x}}},{{\mathbf{y}}}}
Pr\left\{{{\mathbf{x}}},{{\mathbf{y}}}|{{\mathbf{u}}_x},{{\mathbf{u}}_y}
\right\}
    \end{array}
 \label{eq3AAprev}
\end{equation}
%
where $\tilde{{\mathbf{x}}}$ and $\tilde{{\mathbf{y}}}$ are the
jointly estimated information sequences.

Although its optimality, such a joint decoding scheme requires in
general a huge computational effort to be implemented. As a matter
of fact, it requires a squared number of operation per seconds with
respect to unjoint decoding. Such an implementation complexity is
expected in many cases to be too high, particularly when wireless
sensor networks' applications are of concern. In order to get a
simplified receiver structure, let's now observe that by using the
Bayes rule equation (\ref{eq3AAprev}) can be rewritten as:
\begin{equation}
    \begin{array}{cl}
    \left\{\tilde{{\mathbf{x}}},\tilde{{\mathbf{y}}}\right\} = \mathop{arg ~
max}\limits_{{{\mathbf{x}}},{{\mathbf{y}}}}
Pr\left\{{\mathbf{x}}|{\mathbf{y}},{{\mathbf{u}}_x},{{\mathbf{u}}_y}\right\}Pr\left\{{\mathbf{y}}|{{\mathbf{u}}_x},{{\mathbf{u}}_y}\right\}
    \end{array}
 \label{eq3AAprev1}
\end{equation}
The above expression can be simplified by observing that
${{\mathbf{u}}_y}$ is e noisy version of ${\mathbf{y}}$ and that the
noise is independent of ${\mathbf{x}}$. Hence, (\ref{eq3AAprev1})
can be rewritten as:
\begin{equation}
    \begin{array}{cl}
    \left\{\tilde{{\mathbf{x}}},\tilde{{\mathbf{y}}}\right\} = \mathop{arg ~
max}\limits_{{{\mathbf{x}}},{{\mathbf{y}}}}
Pr\left\{{\mathbf{x}}|{\mathbf{y}},{{\mathbf{u}}_x}\right\}Pr\left\{{\mathbf{y}}|{{\mathbf{u}}_x},{{\mathbf{u}}_y}\right\}
    \end{array}
 \label{eq3AAprev2}
\end{equation}
By making similar considerations as above, it is straightforward to
derive from (\ref{eq3AAprev2}) the equivalent decoding rule:
\begin{equation}
    \begin{array}{cl}
    \left\{\tilde{{\mathbf{x}}},\tilde{{\mathbf{y}}}\right\} = \mathop{arg ~
max}\limits_{{{\mathbf{x}}},{{\mathbf{y}}}}
Pr\left\{{\mathbf{y}}|{\mathbf{x}},{{\mathbf{u}}_y}\right\}Pr\left\{{\mathbf{x}}|{{\mathbf{u}}_x},{{\mathbf{u}}_y}\right\}
    \end{array}
 \label{eq3AAprev3}
\end{equation}
Let's now consider the following system of equations:
\begin{equation}
    \begin{array}{cl}
    \tilde{{\mathbf{x}}} = \mathop{arg ~
max}\limits_{{{\mathbf{x}}}}
Pr\left\{{\mathbf{x}}|{\tilde{\mathbf{y}}},{{\mathbf{u}}_x}\right\}Pr\left\{{\tilde{\mathbf{y}}}|{{\mathbf{u}}_x},{{\mathbf{u}}_y}\right\}\\
    \tilde{{\mathbf{y}}} = \mathop{arg ~
max}\limits_{{{\mathbf{y}}}}
Pr\left\{{\mathbf{y}}|\tilde{{\mathbf{x}}},{{\mathbf{u}}_y}\right\}Pr\left\{\tilde{{\mathbf{x}}}|{{\mathbf{u}}_x},{{\mathbf{u}}_y}\right\}
    \end{array}
 \label{eq3AAprev4}
\end{equation}
It is straightforward to observe that the above system has at least
one solution, that is the optimum MAP
solution given by (\ref{eq3AAprev2}) or (\ref{eq3AAprev3}). \\
It is also worth noting that
$Pr\left\{{\tilde{\mathbf{y}}}|{{\mathbf{u}}_x},{{\mathbf{u}}_y}\right\}$
and
$Pr\left\{\tilde{{\mathbf{x}}}|{{\mathbf{u}}_x},{{\mathbf{u}}_y}\right\}$
are constant terms in (\ref{eq3AAprev4}). Therefore, the decoding
problem (\ref{eq3AAprev4}) can be rewritten as:
\begin{equation}
    \begin{array}{cl}
    \tilde{{\mathbf{x}}} = \mathop{arg ~
max}\limits_{{{\mathbf{x}}}}
Pr\left\{{\mathbf{x}}|{\tilde{\mathbf{y}}},{{\mathbf{u}}_x}\right\}\\
    \tilde{{\mathbf{y}}} = \mathop{arg ~
max}\limits_{{{\mathbf{y}}}}
Pr\left\{{\mathbf{y}}|\tilde{{\mathbf{x}}},{{\mathbf{u}}_y}\right\}
    \end{array}
 \label{eq3AAprev5}
\end{equation}
In (\ref{eq3AAprev5}) the decoding problem has been split into two
sub-problems: in each sub-problem the decoder detects one
information signal basing on a-priori information given by the other
decoder. A-priori information will be referred to as
side-information in the following.\\
A solution of the above problem could be obtained by means of an
iterative approach, thus noticeably reducing the implementation
complexity with respect to optimum joint decoding. However,
demonstrating if the iterative decoding scheme converges and, if it
does, to which kind of solution it converges, is a very cumbersome
problem which is out of the scope of this paper. As in the
traditional turbo decoding problem, we are instead interested in
deriving a practical method to solve (\ref{eq3AAprev5}).

\begin{figure}
\begin{center}
\includegraphics[width=0.7\textwidth]{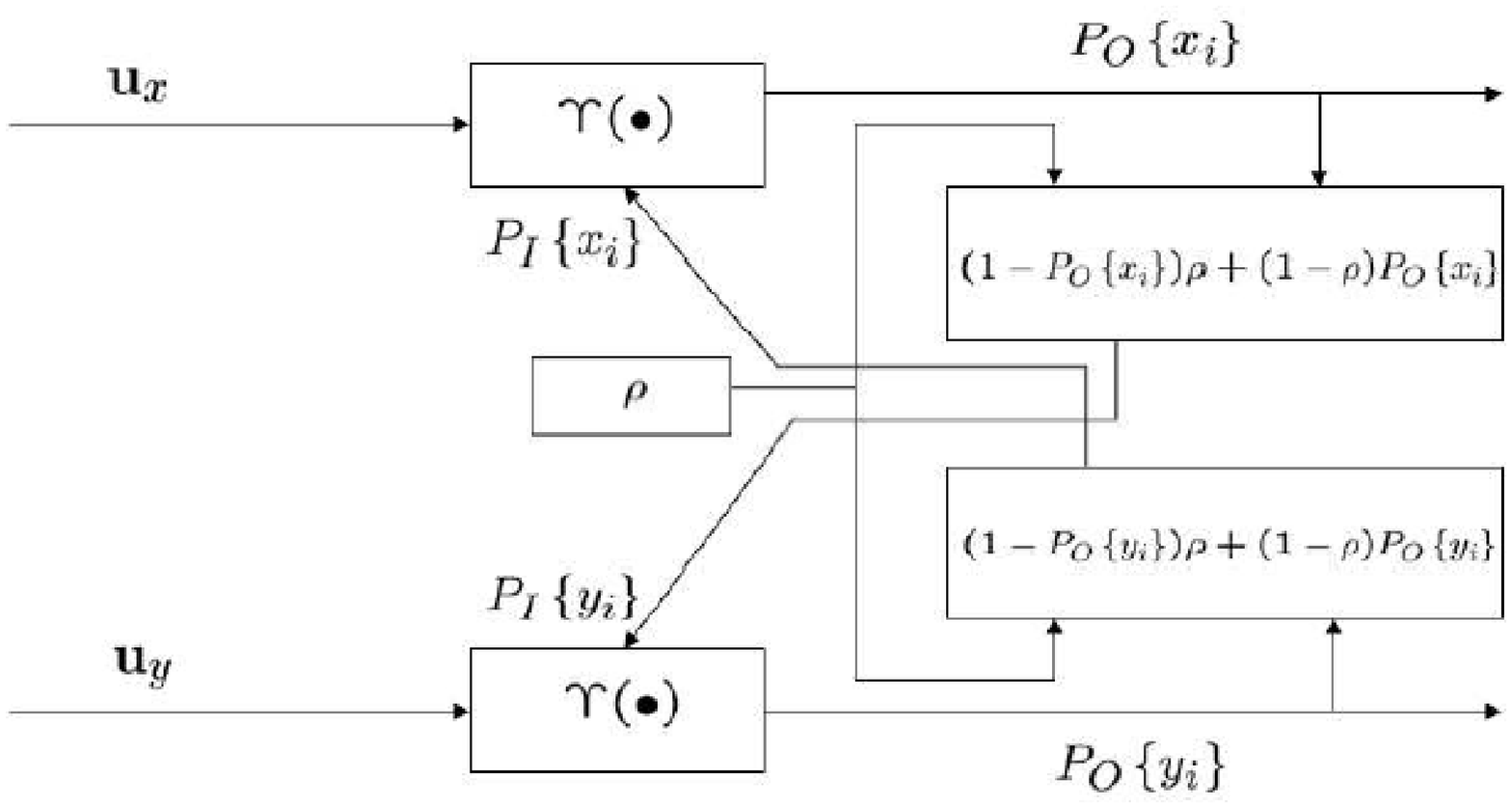}
\end{center} \caption{SOVA Iterative decoding scheme} \label{Fig2i}
\end{figure}

To this aim, classical Soft Input Soft Output (SISO) decoding
schemes, where the decoder gets at its input a-priori information of
input bits and produce at its output a MAP estimation of the same
bits, can be straightforwardly used in this scenario. MAP
estimations and a-priori information are often expressed as
log-likelihood probabilities ratios, which can be easily converted
in bit probabilities \cite{Sklar}. Let denote by
$P_I\left\{x_i\right\}$ and $P_I\left\{y_i\right\}$ the a-priori
probabilities at the SISO decoders' inputs, and by
$P_O\left\{x_i\right\}$ and $P_O\left\{y_i\right\}$  the
a-posteriori probabilities evaluated by the two decoders. In order
to let the iterative scheme working, it is necessary to convert
a-posteriori probabilities evaluated at $j-th$ step into a-priori
probabilities for the $(j+1)-th$ step. According to the correlation
model between the information signals, we get:
\begin{equation}
    \begin{array}{cl}
P_I\left\{y_i\right\}=P_O\left\{x_i\right\}\times\rho +
\left(1-P_O\left\{x_i\right\}\right)\times\left(1-\rho\right)\\
P_I\left\{x_i\right\}=P_O\left\{y_i\right\}\times\rho +
\left(1-P_O\left\{y_i\right\}\right)\times\left(1-\rho\right)
    \end{array}
 \label{eq3DEC1}
\end{equation}
As for the decoding scheme, we consider the Soft Output Viterbi
Decoding (SOVA) scheme depicted in \cite{Sklar}. Denoting to as
$\Upsilon$ the SOVA decoding function, the overall iterative
procedure can be summarized as:
\begin{equation}
    \begin{array}{c}
P^{(1)}_I\left\{x_i\right\} = 0.5; \hfill\\
for ~ j ~ = ~ 1,N \hfill \\
~ ~ ~ ~  P^{(j)}_O\left\{x_i\right\}=\Upsilon\left(P^{(j)}_I\left\{x_i\right\},{{\mathbf{u}}_x}\right); \hfill \\
~ ~ ~ ~
P^{(j)}_I\left\{y_i\right\}=P^{(j)}_O\left\{x_i\right\}\times\rho +
\left(1-P^{(j)}_O\left\{x_i\right\}\right)\times\left(1-\rho\right); \hfill \\
~ ~ ~ ~  P^{(j)}_O\left\{y_i\right\}=\Upsilon\left(P^{(j)}_I\left\{y_i\right\},{{\mathbf{u}}_y}\right); \hfill\\
~ ~ ~ ~
P^{(j)}_I\left\{x_i\right\}=P^{(j)}_O\left\{y_i\right\}\times\rho +
\left(1-P^{(j)}_O\left\{y_i\right\}\right)\times\left(1-\rho\right); \hfill\\
end; \hfill
    \end{array}
 \label{eq3DEC2}
\end{equation}
where $N$ is the number of iterations. In Figure \ref{Fig2i} the
iterative SOVA joint decoding scheme described above is depicted. We
assume that the correlation factor $\rho$ between the information
signals is perfectly known/estimated at the receiver. Such an
assumption is reasonable since $\rho$ is expected to remain almost
constant for long time.

\section{Pairwise error probability}

We now are interested in evaluating the performance of the iterative
joint-decoding scheme. To this aim, we consider a simplified problem
where the side-information provided to the other decoder is without
errors, i.e., it is equal to the original information signal.
Without loss of generality, let focus on the first decoder:
\begin{equation}
    \begin{array}{cl}
    \tilde{{\mathbf{x}}} = \mathop{arg ~
max}\limits_{{{\mathbf{x}}}}
Pr\left\{{\mathbf{x}}|\hat{{{\mathbf{y}}}},{{\mathbf{u}}_x}\right\}
    \end{array}
 \label{eq3AAprev6}
\end{equation}
where $\hat{{{\mathbf{y}}}}$ is the information signal which has
been actually acquired by the second sensor. On account of the ideal
side-information assumption, $\hat{{{\mathbf{y}}}}$ is correlated
with ${\mathbf{x}}$ according to the model $Pr\left\{x_i =
\hat{{{{y}}}}_i\right\} = \rho$. To get an insight into how the
ideal side-information assumption may affect the decoder's
performance, let's start by denoting to as $e_s =
\hat{{{\mathbf{x}}}} \oplus \hat{{{\mathbf{y}}}}$ the information
signals' cross-error profile, $\hat{{{\mathbf{x}}}}$ being the
information signal which has been actually transmitted by the first
transmitter. Moreover, let denote to as $e_d =
\tilde{{{\mathbf{y}}}} \oplus \hat{{{\mathbf{y}}}}$ the error
profile of the second decoder after decoding (\ref{eq3AAprev5}). If
we make the reasonable assumption that $e_s$ and $e_d$ are
independent, the actual side-information $\tilde{{{\mathbf{y}}}}$ is
correlated with ${\mathbf{x}}$ according to the model $Pr\left\{x_i
= \tilde{{{{y}}}}_i\right\} = \rho' \le \rho$, where:
\begin{equation}
    \begin{array}{cl}
        \rho'= \rho \times (1-P_b) + (1-\rho)\times P_b
            \end{array}
 \label{eq3AAprev6bis}
\end{equation}
and $P_b =Pr\left\{\tilde{{{{y}}}}_i \ne \hat{{{{y}}}}_i\right\}$ is
the bit error probability. It is clear from the above expression
that for small $P_b$ we get $\rho' \cong \rho$, i.e., we expect that
for low bit error probability, the ideal side-information assumption
leads to an accurate performance evaluation of the iterative
decoding (\ref{eq3AAprev5}).
This expectation will be confirmed by comparisons with simulation
results in Section V.


By using the Bayes rule and by putting away the constant terms
(i.e., the terms which do not depend on ${{\mathbf{x}}}$), it is now
straightforward to get from (\ref{eq3AAprev6}) the equivalent
decoding rule:
\begin{equation}
    \begin{array}{cl}
    \tilde{{\mathbf{x}}} = \mathop{arg ~
max}\limits_{{{\mathbf{x}}}}
Pr\left\{{{\mathbf{u}}_x}|{\mathbf{x}}\right\}Pr\left\{{{\mathbf{x}}}|\hat{{\mathbf{y}}}\right\}
    \end{array}
 \label{eq3AAprev6bis}
\end{equation}
Substituting for ${{\mathbf{u}}_x}$ the expression given in
(\ref{eq2AA}) and considering the AWGN channel model proposed in the
previous Section, (\ref{eq3AAprev6bis}) can be rewritten as:

%
%

\begin{equation}
    \begin{array}{cl}
    \tilde{{\mathbf{x}}} = \mathop{arg ~
max}\limits_{{{\mathbf{x}}}} \left[\sqrt{2
\xi_{c,x}}\sum\limits_{i=0}^{n-1} u_{i,x}z_i + N_0 \times
ln\left(Pr\left\{{{\mathbf{x}}}|\hat{{\mathbf{y}}}\right\}\right)\right]
    \end{array}
 \label{eq4AA}
\end{equation}
Let's now denote by ${{\mathbf{x}}_t}$ the transmitted information
signal, and by ${{\mathbf{x}}_e} \neq {{\mathbf{x}}_t}$ the
estimated sequence. Moreover, let's denote by ${{\mathbf{z}}_e} \neq
{{\mathbf{z}}_t}$ the corresponding codewords and by $\gamma_{b,x} =
\frac{\xi_{b,x}}{N_0}$. Conditioning to $\hat{{\mathbf{y}}}$, the
pairwise error probability for a given $\gamma_{b,x}$ can be defined
as the probability that the metric (\ref{eq4AA}) evaluated for
${\mathbf{z}} = {{\mathbf{z}}_e}$ and ${\mathbf{x}} =
{{\mathbf{x}}_e}$ is higher than that evaluated for ${\mathbf{z}} =
{{\mathbf{z}}_t}$ and ${\mathbf{x}} = {{\mathbf{x}}_t}$. Such a
probability can be expressed as:
\begin{equation}
    \begin{array}{cl}
    P_e\left({{\mathbf{x}}_t},{{\mathbf{x}}_e},\gamma_{b,x}|\hat{{\mathbf{y}}}\right) =
    \Pr\left\{\sqrt{2
\xi_{c,x}}\sum\limits_{i=0}^{n-1}
u_{i,x}\left(z_{i,e}-z_{i,t}\right)
 - N_0 \times
ln\left(\frac{Pr\left\{{{\mathbf{x}}_t}|\hat{{\mathbf{y}}}\right\}}{Pr\left\{{{\mathbf{x}}_e}|\hat{{\mathbf{y}}}\right\}}\right)
> 0\right\}
    \end{array}
 \label{eq5AA}
\end{equation}
Let's now introduce the hamming distance $d_z =
D\left({{\mathbf{z}}_t},{{\mathbf{z}}_e}\right)$ between the
transmitted and the estimated codewords. Substituting for
${\mathbf{u}}_x$ in (\ref{eq5AA}) the expression given in
(\ref{eq2AA}), it is straightforward to obtain:
\begin{equation}
    \begin{array}{cl}
    P_e\left({{\mathbf{x}}_t},{{\mathbf{x}}_e},\gamma_{b,x}|\hat{{\mathbf{y}}}\right) =
    0.5 erfc\left[\sqrt{r d_z \gamma_{b,x}}  +\frac{1}{4\sqrt{r d_z \gamma_{b,x} }}ln\left(\frac{Pr\left\{{{\mathbf{x}}_t}|\hat{{\mathbf{y}}}\right\}}{Pr\left\{{{\mathbf{x}}_e}|\hat{{\mathbf{y}}}\right\}}\right)
    \right]
    \end{array}
 \label{eq6AA}
\end{equation}
where $\gamma_{b,x} = \frac{\xi_{b,x}}{N_0}$ and $erfc$ is the
complementary error function. Notice that the term in (\ref{eq6AA})
which takes into account the side-information $\hat{{\mathbf{y}}}$
is given by the natural logarithm of a ratio of probabilities. It is
straightforward to note that such a term can be positive or
negative, depending wether the Hamming distance
$D\left({{\mathbf{x}}_t},\hat{{\mathbf{y}}}\right)$ is higher or
lower than $D\left({{\mathbf{x}}_e},\hat{{\mathbf{y}}}\right)$. Of
course, for high $\rho$, the probability that such term becomes
negative is low, and hence one expects that on the average the
effect of a-priori information is positive, i.e., it increases the
argument of the erfc function or, equivalently, it reduces the
pairwise error probability. To elaborate, let's now introduce:
\begin{equation}
    \begin{array}{cl}
    \Gamma_{i,t} = x_{i,t} \oplus \hat{y}_{i}\\
    \Gamma_{i,e} = x_{i,e} \oplus \hat{y}_{i}
    \end{array}
 \label{eq7AA}
\end{equation}
where $\oplus$ is the XOR operator. Hence, it can be easily derived:
\begin{equation}
    \begin{array}{cl}
    \frac{Pr\left\{{{\mathbf{x}}_t}|\hat{{\mathbf{y}}}\right\}}{Pr\left\{{{\mathbf{x}}_e}|\hat{{\mathbf{y}}}\right\}}
    =
    \frac{\prod\limits_{i=0}^{k-1}\rho^{1-\Gamma_{i,t}}(1-\rho)^{\Gamma_{i,t}}}{\prod\limits_{i=0}^{k-1}\rho^{1-\Gamma_{i,e}}(1-\rho)^{\Gamma_{i,e}}}
    =\prod\limits_{i=0}^{k-1}\rho^{\Gamma_{i,e}-\Gamma_{i,t}} \times (1-\rho)^{\Gamma_{i,t}-\Gamma_{i,e}}
    \end{array}
 \label{eq8AA}
\end{equation}
The above expression can be further simplified by observing that
$\Gamma_{i,t}-\Gamma_{i,e}$ is different from zero only for $x_{i,t}
\oplus x_{i,e} = 1$. Hence, by introducing the set $I = \left\{i:
x_{i,t} \oplus x_{i,e} = 1\right\}$, equation (\ref{eq6AA}) can be
rewritten:
\begin{equation}
    \begin{array}{cl}
    P_e\left({{\mathbf{x}}_t},{{\mathbf{x}}_e},\gamma_{b,x}|\hat{{\mathbf{y}}}\right) =
    0.5 erfc\left[{\sqrt{r d_z \gamma_{b,x} }} + \frac{1}{4\sqrt{r d_z \gamma_{b,x} }}ln\left(\prod\limits_{i \in I}\rho^{\Gamma_{i,e}-\Gamma_{i,t}} \times (1-\rho)^{\Gamma_{i,t}-\Gamma_{i,e}}\right)
    \right]
    \end{array}
 \label{eq10AA}
\end{equation}
Let's introduce the term $d_x$ as the Hamming distance between the
transmitted and the estimated information signals, i.e., $d_x =
\sum\limits_{i = 0}^{k-1} x_{i,t} \oplus x_{i,e}$. Notice that $d_x$
is the dimension of the set $I$ and, hence, the product over $I$ in
(\ref{eq10AA}) is a product of $d_x$ terms.\\
The problem of evaluating the pairwise error probability in presence
of a-priori soft information has already been derived in a previous
work \cite{Hagenauer} and cited in a recent work \cite{Mondinf}. In
\cite{Hagenauer} and \cite{Mondinf} the a-priori information is
expressed as log-likelihood value of the information signal and is
referred to as $L$ (e.g., see equation (5) of \cite{Mondinf}).
Notice that, according to the notations of this paper, such a
log-likelihood information can be expressed as $L =
ln\left(\frac{\rho}{1-\rho}\right)$. Note also that in equation (5)
of \cite{Mondinf} the pairwise error probability is expressed as
$P_d = \frac{1}{2}erfc\left(\sqrt{\frac{r d
E_b}{N_0}\left(1+\frac{w_d}{m_d}\frac{L}{4 r d
E_b/N_0}\right)^2}\right)$, that, through easy mathematics, becomes
$P_d = \frac{1}{2}erfc\left(\sqrt{\frac{r d E_b}{N_0}}+\frac{w_d}{
m_d}\frac{L}{4\sqrt{r d E_b/N_0}}\right)$. Hence, in
\cite{Hagenauer} and \cite{Mondinf} the logarithm of the product
over $I$ (\ref{eq10AA}) is set equal to the sum of the a-priori
information log-likelihood values of $x_{i,t}$, i.e., it is set
equal to $\frac{w_d}{m_d} L = d_x L$. Considering the notation of
this paper, this is equivalent to set $\Gamma_{i,e} = 1$ and
$\Gamma_{i,t} = 0$, for $i \in I$, i.e., to assume that there is a
perfect match between the a-priori information $\hat{{\mathbf{y}}}$
and the actually transmitted information $\hat{{\mathbf{x}}}$. This
assumption would lead to heavily underestimate the pairwise error
probability, as it will be shown at the end of this Section.\\
To further elaborate, notice that the terms
$\rho^{\Gamma_{i,e}-\Gamma_{i,t}} \times
(1-\rho)^{\Gamma_{i,t}-\Gamma_{i,e}}$, with $i \in I$, can take the
following
values:\\
I) $\frac{\rho}{1-\rho}$, if $x_{i,t} \oplus \hat{y}_{i} = 0$ \\
II) $\frac{1-\rho}{\rho}$, if $x_{i,t} \oplus \hat{y}_{i} = 1$ \\
Let's now define by $\varepsilon_i = \overline{({x_{i,t} \oplus
\hat{y}_{i}})}$, the logical not of $x_{i,t} \oplus \hat{y}_{i}$.
Then, $P_e$ can be rewritten as:
\begin{equation}
    \begin{array}{cl}
    P_e\left({{\mathbf{x}}_t},{{\mathbf{x}}_e},\gamma_{b,x}|\hat{{\mathbf{y}}}\right) =
    0.5 erfc\left\{\sqrt{r d_z \gamma_{b,x}}  +\frac{1}{4\sqrt{r d_z \gamma_{b,x} }}ln\left[\left(\frac{\rho}{1-\rho}\right)^{\sum\limits_{k = 1}^{d_x}\varepsilon_{i(k)}}\left(\frac{1-\rho}{\rho}\right)^{d_x-\sum\limits_{k = 1}^{d_x}
    \varepsilon_{i(k)}}\right]
    \right\}
    \end{array}
 \label{eq11AA}
\end{equation}
where indexes $i(k)$, $k = 1,\ldots,d_x$ are all the elements of the
set $I$. Note that $P_e$ expressed in (\ref{eq11AA}) is a function
of $\varepsilon_i$, $i \in I$, rather then of the whole vector
${\hat{{\mathbf{y}}}}$. Hence, we can write:
\begin{equation}
    \begin{array}{cl}
    P_e\left({{\mathbf{x}}_t},{{\mathbf{x}}_e},\gamma_{b,x}|\varepsilon_{i(1)},\varepsilon_{i(2)},\ldots,\varepsilon_{i(d_x)}\right) =
    0.5 erfc\left\{\sqrt{r d_z \gamma_{b,x}} + \right. \\ \left. +\frac{1}{4\sqrt{r d_z \gamma_{b,x} }}ln\left[\left(\frac{\rho}{1-\rho}\right)^{\sum\limits_{k = 1}^{d_x}\varepsilon_{i(k)}}\left(\frac{1-\rho}{\rho}\right)^{d_x-\sum\limits_{k = 1}^{d_x}
    \varepsilon_{i(k)}}\right]
    \right\}
    \end{array}
 \label{eq11AAbis}
\end{equation}
%
%
 Notice that $\varepsilon_i$
is by definition equal to one with probability $\rho$ and equal to
zero with probability $1-\rho$. Hence, it is possible to filter out
the dependence on $\varepsilon_i$ in (\ref{eq11AA}), thus obtaining
an average pairwise error probability given by:
\begin{equation}
    \begin{array}{cl}
    {P}_e\left({{\mathbf{x}}_t},{{\mathbf{x}}_e},\gamma_{b,x}\right) =
\sum\limits_{\varepsilon_{i(1)}=\{0,1\}} \ldots
\sum\limits_{\varepsilon_{i(d_x)}=\{0,1\}}
P_e\left({{\mathbf{x}}_t},{{\mathbf{x}}_e},\gamma_{b,x}|\varepsilon_{i(1)},\ldots,\varepsilon_{i(d_x)}\right)
\times \\ \times \rho^{\sum\limits_{k =
1}^{d_x}\varepsilon_{i(k)}}(1-\rho)^{d_x-\sum\limits_{k = 1}^{d_x}
    \varepsilon_{i(k)}}
    \end{array}
 \label{eq12AA}
\end{equation}

It is now convenient for our purposes to observe from
(\ref{eq11AAbis}) and (\ref{eq12AA}) that the pairwise error
probability can be extensively expressed as a function of solely the
hamming distances $d_z$ and $d_x$ as:
\begin{equation}
    \begin{array}{cl}
    {P}_e\left(d_z,d_x,\gamma_{b,x}\right)
= \sum\limits_{\varepsilon_{i(1)}=\{0,1\}} \ldots
\sum\limits_{\varepsilon_{i(d_x)}=\{0,1\}}
   0.5 erfc\left\{\sqrt{r d_z \gamma_{b,x}} + \right. \\ \left. +\frac{1}{4\sqrt{r d_z \gamma_{b,x} }}ln\left[\left(\frac{\rho}{1-\rho}\right)^{\sum\limits_{k = 1}^{d_x}\varepsilon_{i(k)}}\left(\frac{1-\rho}{\rho}\right)^{d_x-\sum\limits_{k = 1}^{d_x}
    \varepsilon_{i(k)}}\right]
    \right\} \times \rho^{\sum\limits_{k =
1}^{d_x}\varepsilon_{i(k)}}(1-\rho)^{d_x-\sum\limits_{k = 1}^{d_x}
    \varepsilon_{i(k)}}
    \end{array}
 \label{eq12AAbis}
\end{equation}

\begin{figure}
\begin{center}
\includegraphics[width=0.8\textwidth]{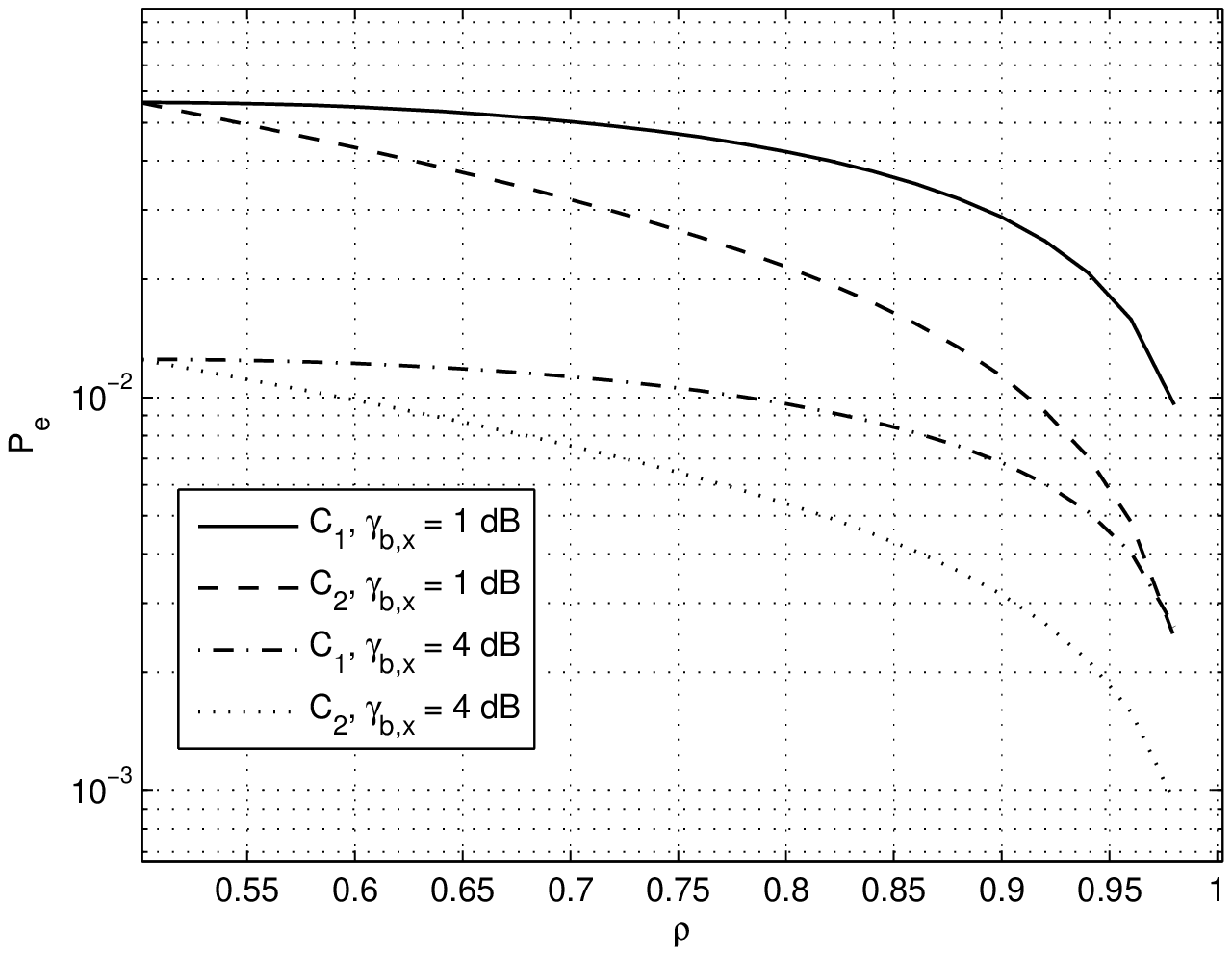}
\end{center} \caption{Bit error probability curves in the uncoded case} \label{Fig3f}
\end{figure}

Equation (\ref{eq12AAbis}) gives rise to interesting considerations
about the properties of good channel codes. In particular, let's
observe that the term $\sum\limits_{k = 1}^{d_x} \varepsilon_{i(k)}$
plays a fundamental role in determining the pairwise error
probability. Indeed, making the natural assumption $\rho
> 0.5$, if $\sum\limits_{k = 1}^{d_x}
\varepsilon_{i(k)} \le \lfloor d_x/2 \rfloor$ the argument of the
logarithm is less than one, and, hence, the performance is affected
by signal-to-noise-ratio reduction (the argument of the $erfc$
function diminishes). Note that, the lowest $\sum\limits_{k =
1}^{d_x} \varepsilon_{i(k)}$ the highest the performance
degradation. Hence, it is important that such bad situations occur
with low probability. On the other hand, the highest $d_x$, the
lowest the probability of bad events which is mainly given by the
term $(1-\rho)^{d_x-\sum\limits_{k = 1}^{d_x}\varepsilon_i(k)}$.
Hence, it is expected that a good code design should lead to
associate high Hamming weight information sequences with low Hamming
weight codewords. To be more specific, if we consider convolutional
codes it is expected that recursive schemes work better than
non-recursive ones. This conjecture will be confirmed in the next
Sections.\\
%
To give a further insight into the analysis derived so far, and to
provide a comparison with the Hagenauer's bounds reported in
\cite{Hagenauer} and \cite{Mondinf}, let's now consider the uncoded
case. In this simple case $r = k = n = 1$, ${{{x}}_t} = {{{z}}_t}$,
${{{x}}_e} = {{{z}}_e}$ (we have mono-dimensional signals), and $d_x
= d_z = 1$. Moreover, the pairwise error probability becomes the
probability to decode $+1/-1$ when $-1/+1$ has been transmitted,
i.e., it is equivalent to the bit error probability. Without loss of
generality, we assume that the side-information is $\hat{{y}} = 1$,
so that we can denote by $L(x) = ln\left(\frac{\rho}{1-\rho}\right)$
the log-likelihood value of a-priori information for the decoder. It
is straightforward to get from (\ref{eq12AAbis}):
\begin{equation}
    \begin{array}{cl}
    {P}_e\left(\gamma_{b,x}\right)
=  0.5 erfc\left(\sqrt{\gamma_{b,x}} +
\frac{L(x)}{4\sqrt{\gamma_{b,x} }}\right) \times \rho + 0.5
erfc\left(\sqrt{\gamma_{b,x}} - \frac{L(x)}{4\sqrt{\gamma_{b,x}
}}\right) \times (1-\rho)
    \end{array}
 \label{eq12AAtris}
\end{equation}
By following the model proposed in \cite{Hagenauer}, we would get:
\begin{equation}
    \begin{array}{cl}
    {P}_e\left(\gamma_{b,x}\right)
=  0.5 erfc\left(\sqrt{\gamma_{b,x}} +
\frac{L(x)}{4\sqrt{\gamma_{b,x} }}\right)
    \end{array}
 \label{eq12AApoker}
\end{equation}
In Fig. \ref{Fig3f} we show the $P_e$ curves as a function of
$\rho$, computed according to (\ref{eq12AAtris}) and
(\ref{eq12AApoker}) and referred to as $C_1$ and $C_2$,
respectively. Two different $\gamma_{b,x}$ values are considered:
$\gamma_{b,x} = 1$ dB and $\gamma_{b,x} = 4$ dB.\\
By running computer simulations we have verified that, as expected,
$C_1$ represents an exact calculation of the bit error probability
(simulation curves perfectly match $C_1$). Accordingly, it is
evident that the approximation (\ref{eq12AApoker}) is not
satisfying. On the other hand, in \cite{Mondinf} the good match
between simulations and theoretical curves is due to the use of
base-10 logarithm instead of the correct natural logarithm. As a
matter of fact, by using the correct calculation of $L(x)$ one would
observe the same kind of underestimation of bit error probability as
shown in Fig. \ref{Fig3f}.

\section{Packet error probability evaluation and Optimal convolutional code searching strategy}

In this Section, and in the rest of the paper, we consider
convolutional coding schemes \cite{Sklar}, \cite{Proakis}. Such
schemes allow an easy coding implementation with very low power and
memory requirements and, hence, they seem to be particularly
suitable for utilization in wireless sensors' networks.
%
Let's now focus on the evaluation of packet error probability at the
decoder in presence of perfect side-information estimation. As in
traditional convolutional coding, it is possible to derive an upper
bound of the bit error probability as the weighted \footnote{The
weights are the information error weights} sum of the pairwise error
probabilities relative to all paths which diverge from the zero
state and marge again after a certain number of transitions
\cite{Sklar}. This is possible because of the linearity of the code
and because the pairwise error probability (\ref{eq12AAbis}) depends
only on input and output weights $d_x$ and $d_z$, and not on the
actual
transmitted sequence.\\
In particular, it is possible to evaluate the input-output transfer
function $T(W,D)$ by means of the state transition relations over
the modified state diagram \cite{Sklar}. The generic form of
$T(W,D)$ is:
\begin{equation}
    \begin{array}{cl}
T(W,D) = \sum\limits_{w,d}^{} \beta_{w,d} W^w D^d
    \end{array}
 \label{eq1BB}
\end{equation}
where $\beta_{w,d}$ denotes the number of paths that start from the
zero state and reemerge with the zero state and that are associated
with an input sequence of weight $w$, and an output sequence of
weight $d$. Accordingly, we can get an upper bound of the bit error
probability of ${\mathbf{x}}$ as:
\begin{equation}
    \begin{array}{cl}
    {P}_{b,x} \le \sum\limits_{w,d} \beta_{w,d}^{(x)} \times w \times {P}_e\left(d,w,\gamma_{b,x}\right)
    \end{array}
 \label{eq13AA}
\end{equation}
where $\beta_{w,d}^{(x)}$ is the $\beta_{w,d}$ term for the first
encoder's code and $P_e(d,w,\gamma_{b,x})$ is the pairwise error
probability (\ref{eq12AAbis}) for $d_z = d$ and $d_x = w$. On
account of the symmetry of the problem (\ref{eq3AAprev4}), the union
bound of the bit error probability of ${\mathbf{y}}$ is:
\begin{equation}
    \begin{array}{cl}
    {P}_{b,y} \le \sum\limits_{w,d} \beta_{w,d}^{(y)} \times w \times {P}_e\left(d,w,\gamma_{b,y}\right)
    \end{array}
 \label{eq13AAbis}
\end{equation}
where $\beta_{w,d}^{(y)}$ is the $\beta_{w,d}$ term for the second
encoder's code and $\gamma_{b,y} = \frac{\xi_{b,y}}{N_0}$.\\
Following a similar procedure, it is then possible to derive the
packet error probabilities. To this aim, let's start by denoting to
as $L_{pkt}$ the packet data length and let's assume that $L_{pkt}$
is much higher than the constraint lengths of the codes (the
assumption is reasonable for the low complexity convolutional codes
that are considered in this paper). In this case, since the
first-error events which contribute with non negligible terms to the
summations (\ref{eq13AA}) and (\ref{eq13AAbis}) have a length of few
times the code's constraint length, we can assume that the number of
first-error events in a packet is equal to $L_{pkt}$ \footnote{In
other terms we neglect the border effect}. Hence, the upper bounds
${P}_{d,x}$ and ${P}_{d,y}$ of the packet error rate can be easily
derived as:
\begin{equation}
    \begin{array}{cl}
    {P}_{d,x} \le \sum\limits_{w,d} \beta_{w,d}^{(x)} \times L_{pkt} \times {P}_e\left(d,w,\gamma_{b,x}\right)
    \\
    {P}_{d,y} \le \sum\limits_{w,d} \beta_{w,d}^{(y)} \times L_{pkt} \times {P}_e\left(d,w,\gamma_{b,y}\right)
    \end{array}
 \label{eq13AAA}
\end{equation}

Basing on the procedure derived above, it is now possible to
implement an exhaustive search over all possible codes' structures
with the aim of finding the optimum code, intended as the code which
minimizes the average packet error rate upper bound ${P}_{d} =
\frac{{P}_{d,x}+{P}_{d,y}}{2}$. We will assume in the following that
sensor 1 and sensor 2 use the same code, and that $k = 1$ and $n =
2$. In this situation, a code is univocally determined by the
generator polynomials $G^{(1)}(D) = g^{(1)}_{\nu} \times
D^{\nu}+g^{(1)}_{\nu-1}
D^{\nu-1}+g^{(1)}_{\nu-2}D^{\nu-2}+\ldots+g^{(1)}_{1}D^{1}+g^{(1)}_{0}$,
$G^{(2)}(D) = g^{(2)}_{\nu} \times D^{\nu}+g^{(2)}_{\nu-1}
D^{\nu-1}+g^{(2)}_{\nu-2}D^{\nu-2}+\ldots+g^{(2)}_{1}D^{1}+g^{(2)}_{0}$
and by the feedback polynomial $H(D) = h_{\nu} \times
D^{\nu}+h_{\nu-1}
D^{\nu-1}+h_{\nu-2}D^{\nu-2}+\ldots+h_{1}D^{1}+h_{0}$, where $\nu$
is the number of shift registers of the code (i.e., the number of
states is $2^\nu$) and $g^{(1)}_{k} = \left\{0,1\right\}$,
$g^{(2)}_{k} = \left\{0,1\right\}$, $h_{k} = \left\{0,1\right\}$.
Hence, the exhaustive search is performed by considering all
possible polynomials, i.e., all $2^{3(\nu+1)}$ possible values of
$G^{(1)}(D)$, $G^{(2)}(D)$, and $H(D)$. It is worth noting that when
$H(D) = 0$ the code is non-recursive while when $H(D) \ne 0$ the
code becomes recursive. Table I shows the optimum code's structure
obtained by exhaustive search for $\gamma_{b,x} = \gamma_{b,y} = 3$
dB and for $\nu = 3$. Three different values of $\rho$, i.e., $\rho
= 0.8$, $\rho = 0.9$ and $\rho = 0.95$, has been considered and
three different codes, namely $C_{80}$, $C_{90}$ and $C_{95}$, have
been correspondingly obtained.

\begin{table}
\begin{center}
\begin{tabular}{|c||c|c|c|c|c|}
\hline & $C_{80}$ : $p = 0.8$ & $C_{90}$ : $p = 0.9$ & $C_{95}$ : $p = 0.95$ \\
\hline\hline $G^{(1)}(D)$ & $D^3+D^2+1$ & $D^3+D+1$ & $D^3+D+1$ \\
\hline $G^{(2)}(D)$ & $D^3+D^2+D+1$ & $D^3+D^2+D+1$ & $D^3+D^2+1$
\\ \hline $H(D)$ & $D^3+D+1$ & $D^3+D^2+1$ & $D^3+D^2+D+1$
\\ \hline
\end{tabular}
~\\ ~\\ ~\\ $\mathbf{Table~I}$: Generator polynomials of the optimum codes ~\\
\end{center}
\end{table}

As it is evident from previous Sections' analysis, the optimum code
structure depends on the signal to noise ratios, i.e., different
values of $\gamma_{b,x}$ and $\gamma_{b,y}$ lead to different
optimum codes. However, by running the optimum code searching
algorithm for a set of different signal to noise ratios, we have
verified that the optimum code's structure remain the same over a
wide range of $\gamma_{b,x}$ and $\gamma_{b,y}$ and, hence, we can
tentatively state that $C_{80}$, $C_{90}$ and $C_{95}$ are the
optimum codes for $\nu$ = 3 and for $\rho = 0.8$, $\rho = 0.9$ and
$\rho =
0.95$.\\

\section{Results and comparisons}

\begin{figure}
\begin{center}
\includegraphics[width=0.8\textwidth]{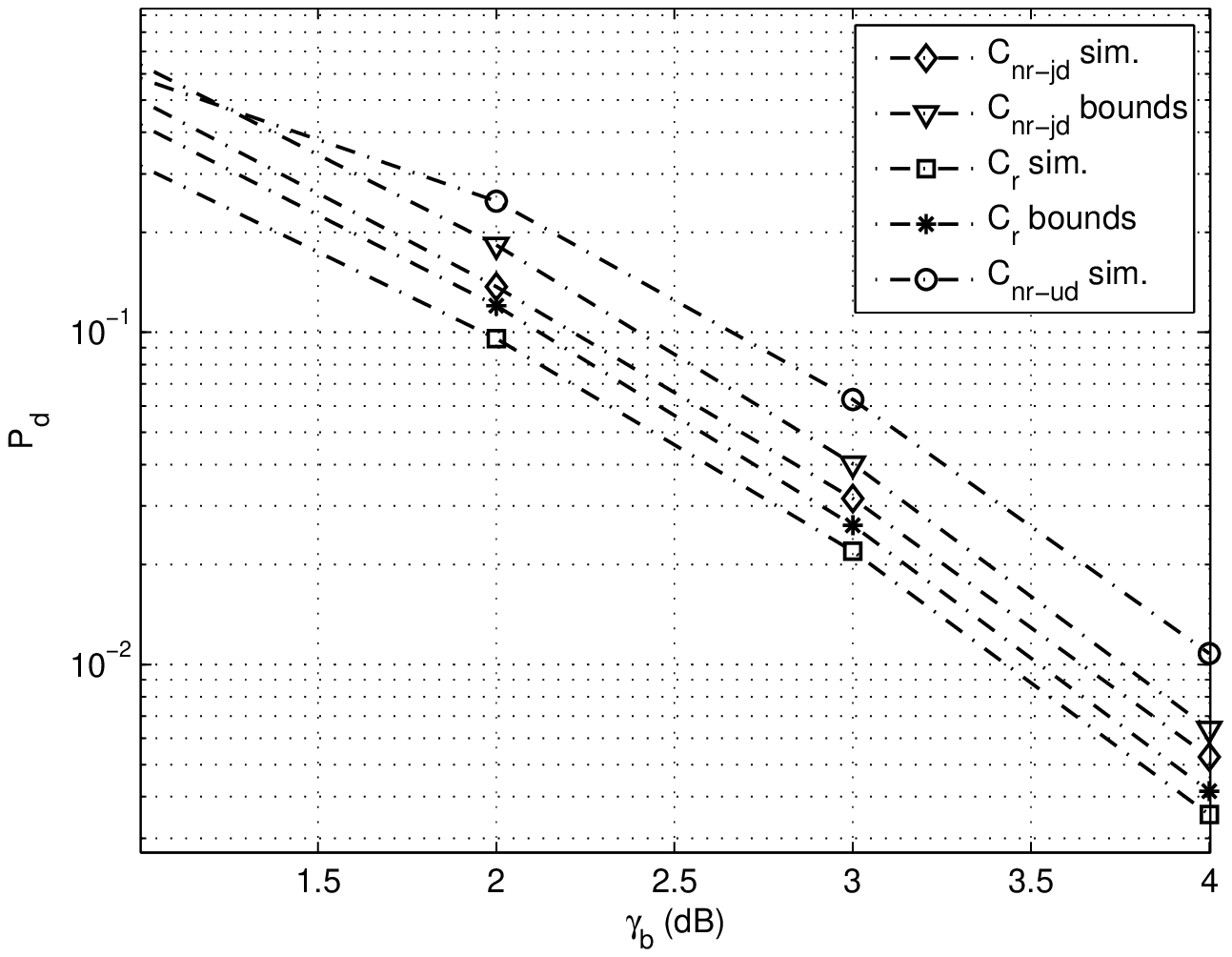}
\end{center} \caption{Simulations results and theoretical bounds for $\rho = 0.8$ and $L_{pkt} = 100$} \label{Fig3}
\end{figure}

\begin{figure}
\begin{center}
\includegraphics[width=0.8\textwidth]{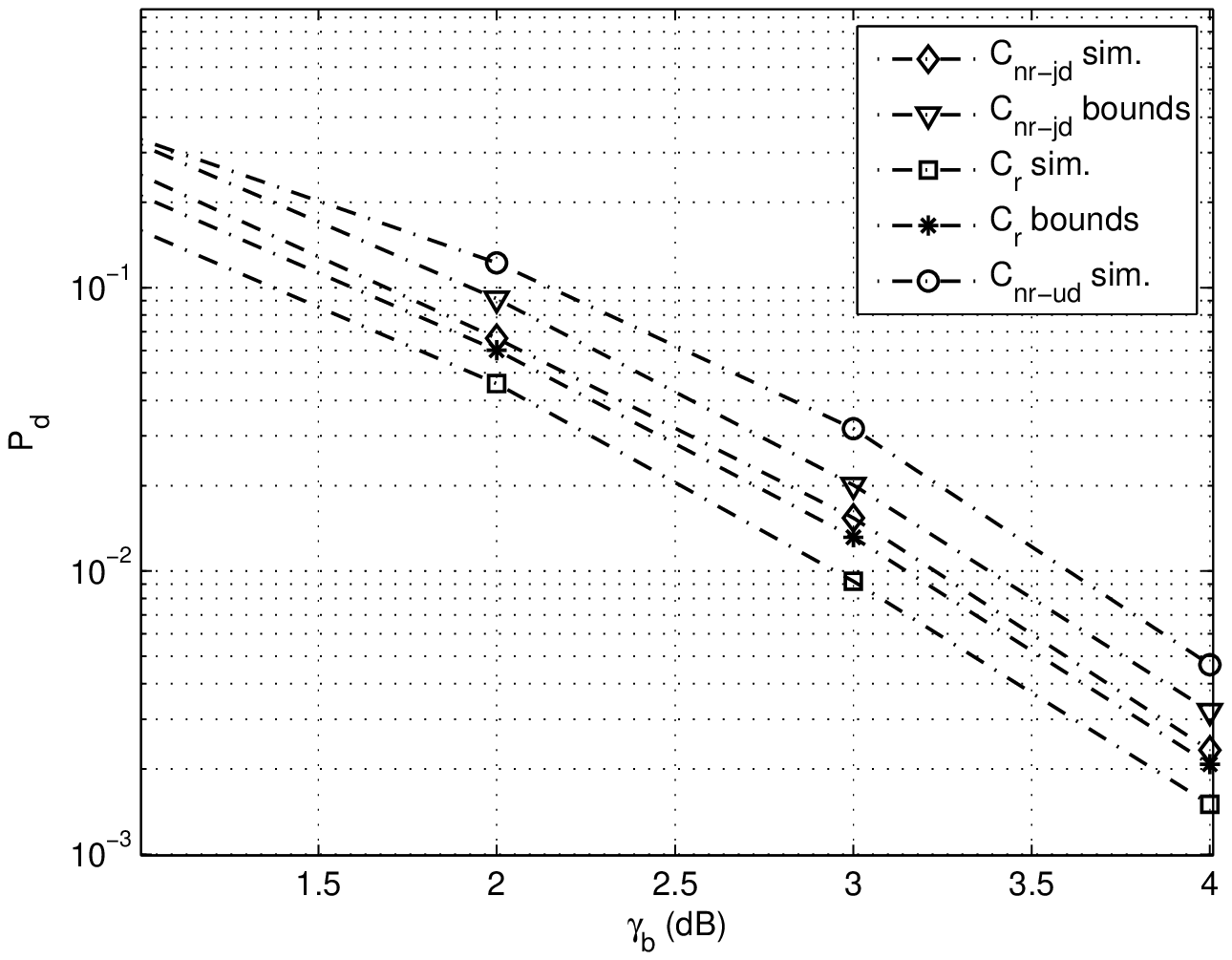}
\end{center} \caption{Simulations results and theoretical bounds for $\rho = 0.8$ and $L_{pkt} = 50$} \label{Fig4}
\end{figure}

\begin{figure}
\begin{center}
\includegraphics[width=0.8\textwidth]{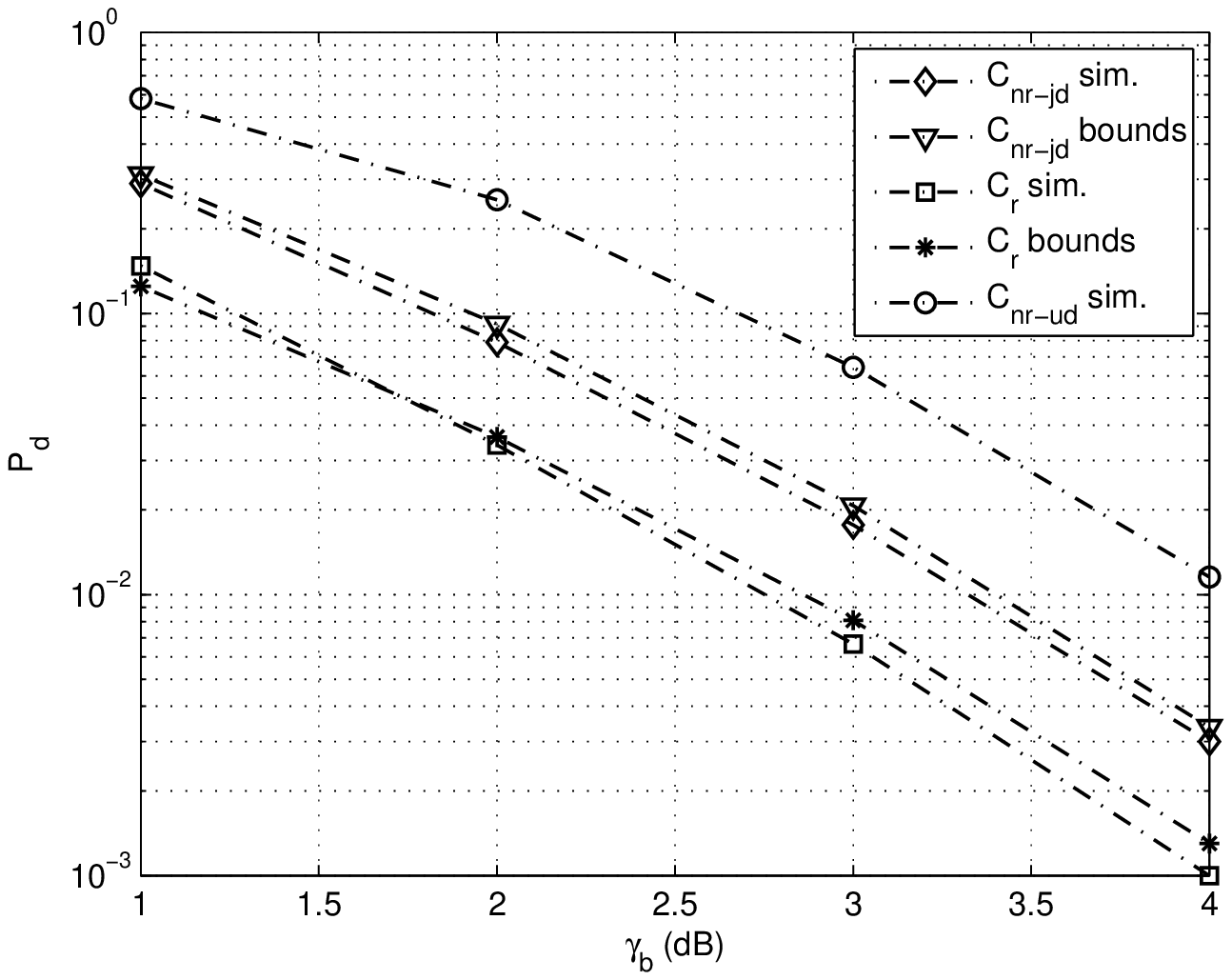}
\end{center} \caption{Simulations results and theoretical bounds for $\rho = 0.9$ and $L_{pkt} = 100$} \label{Fig5}
\end{figure}

\begin{figure}
\begin{center}
\includegraphics[width=0.8\textwidth]{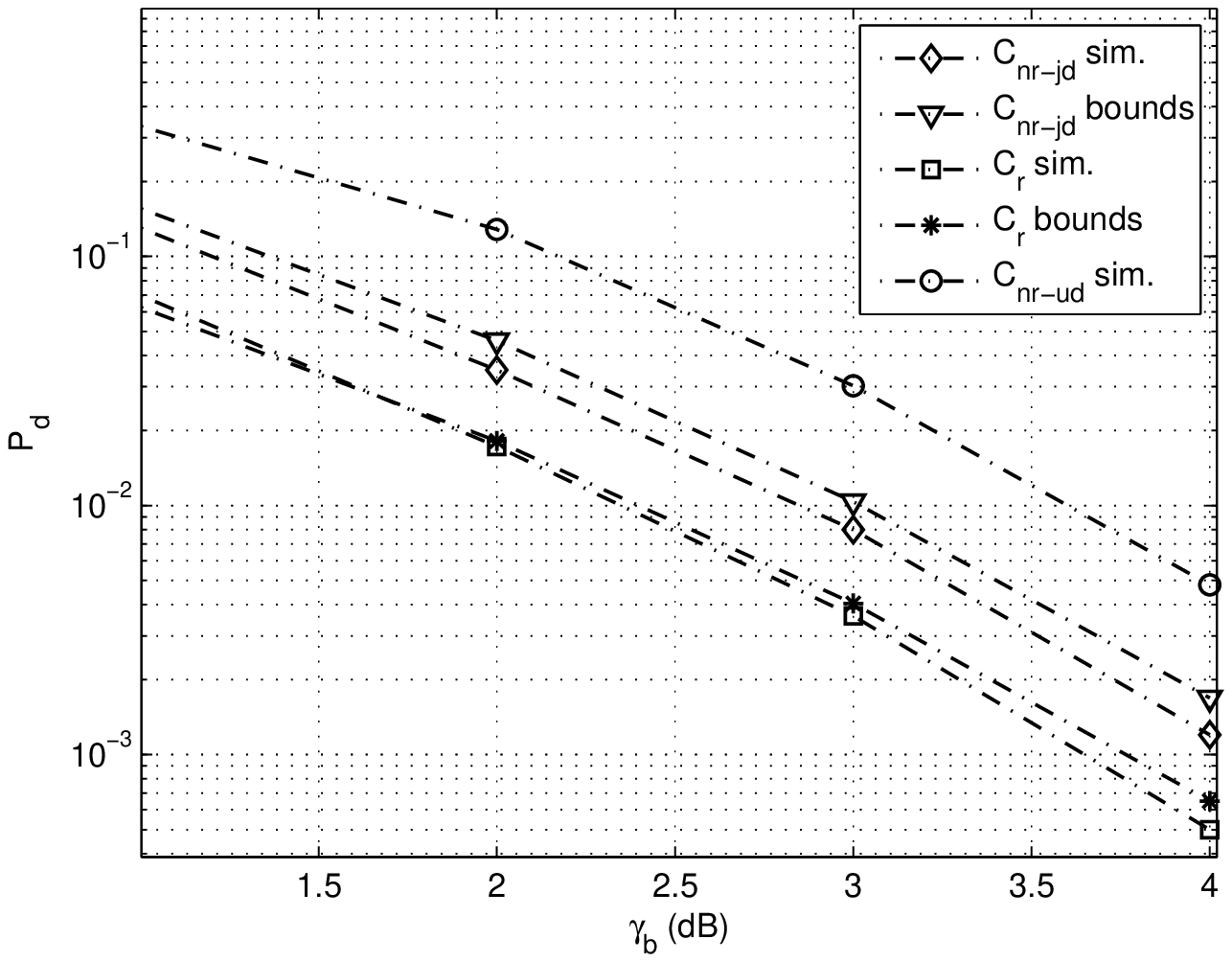}
\end{center} \caption{Simulations results and theoretical bounds for $\rho = 0.9$ and $L_{pkt} = 50$} \label{Fig6}
\end{figure}

\begin{figure}
\begin{center}
\includegraphics[width=0.8\textwidth]{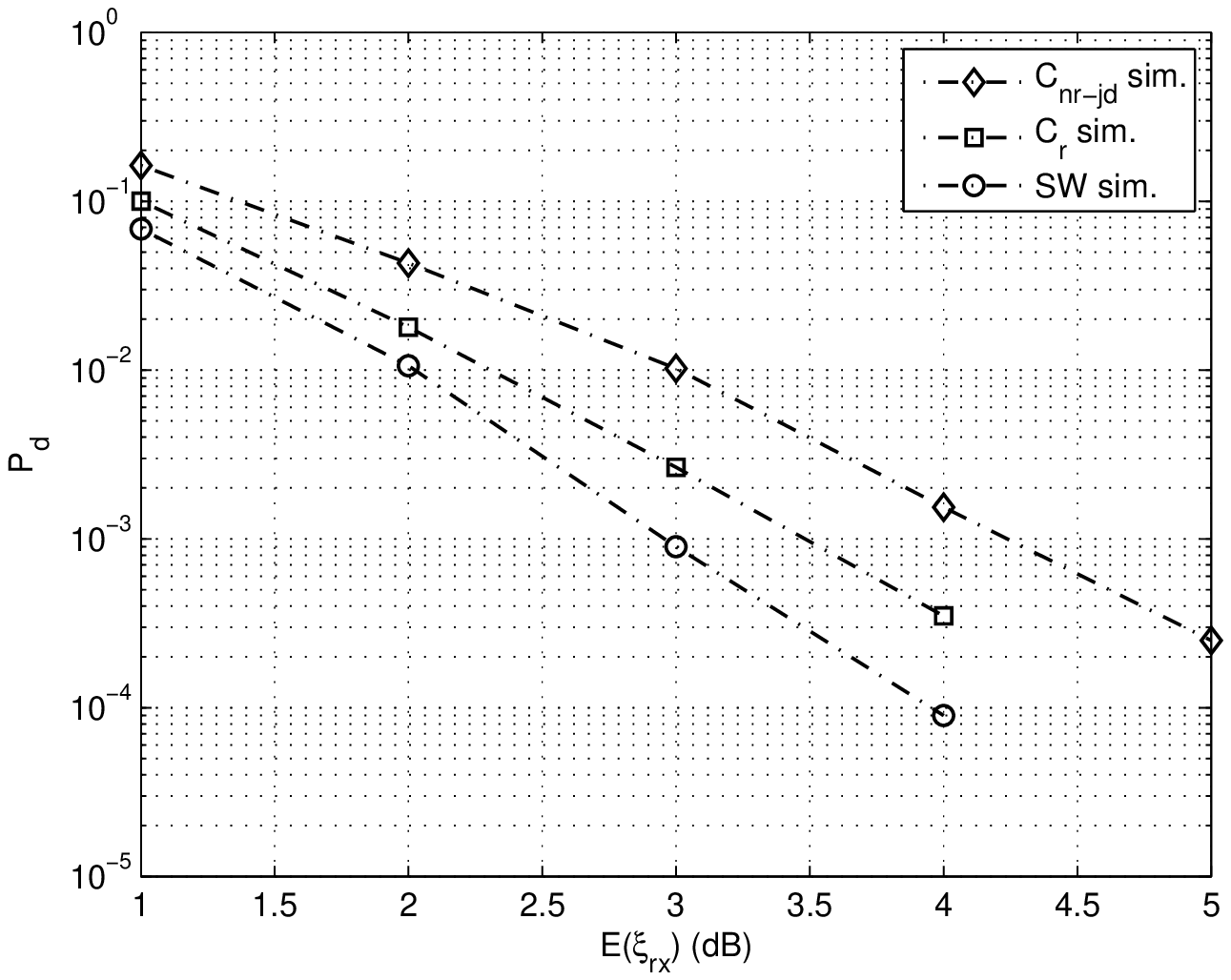}
\end{center} \caption{Comparison with the SW case: AWGN channel} \label{Fig7}
\end{figure}

\begin{figure}
\begin{center}
\includegraphics[width=0.8\textwidth]{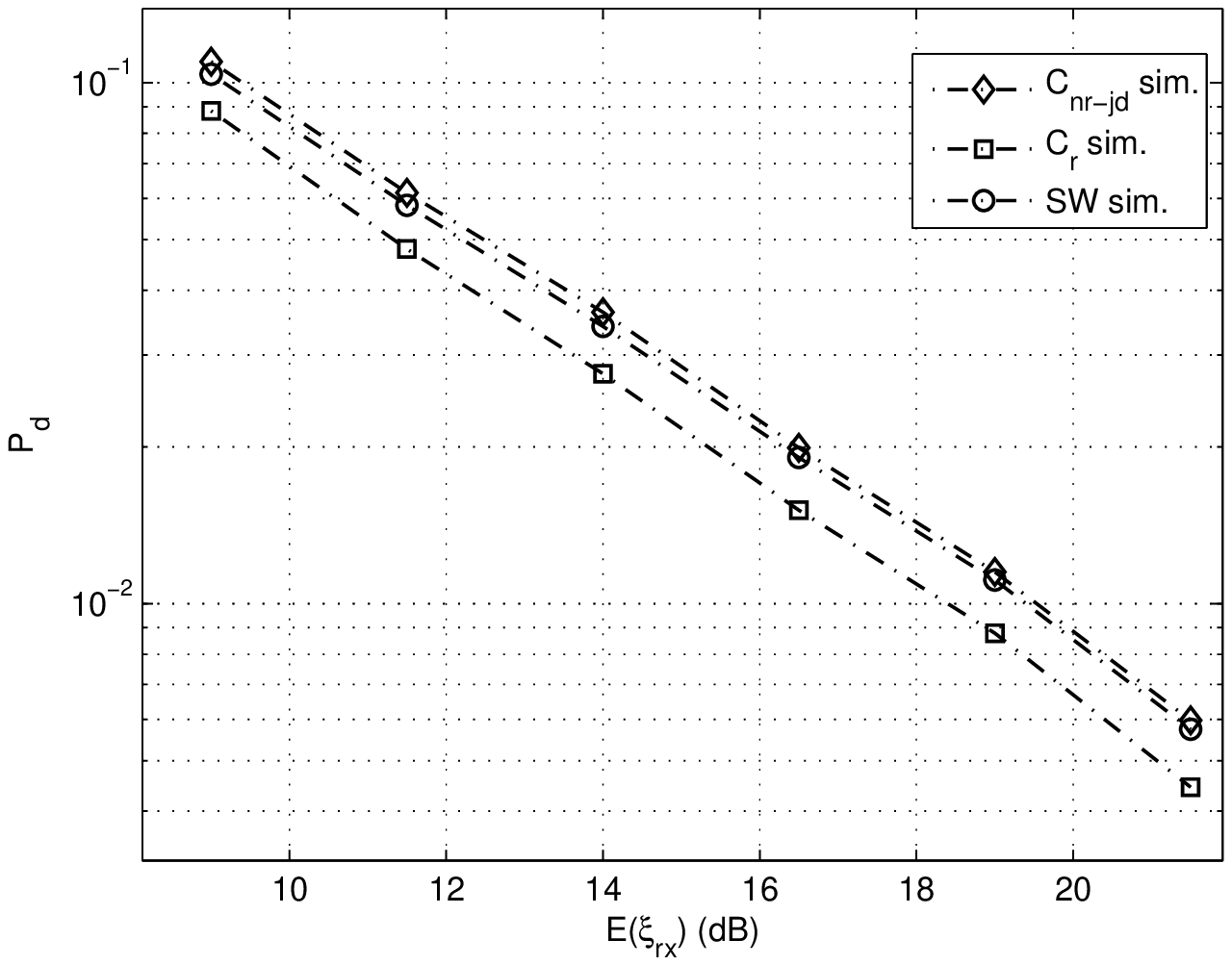}
\end{center} \caption{Comparison with the SW case: Rayleigh channel model} \label{Fig8}
\end{figure}

\begin{figure}
\begin{center}
\includegraphics[width=0.8\textwidth]{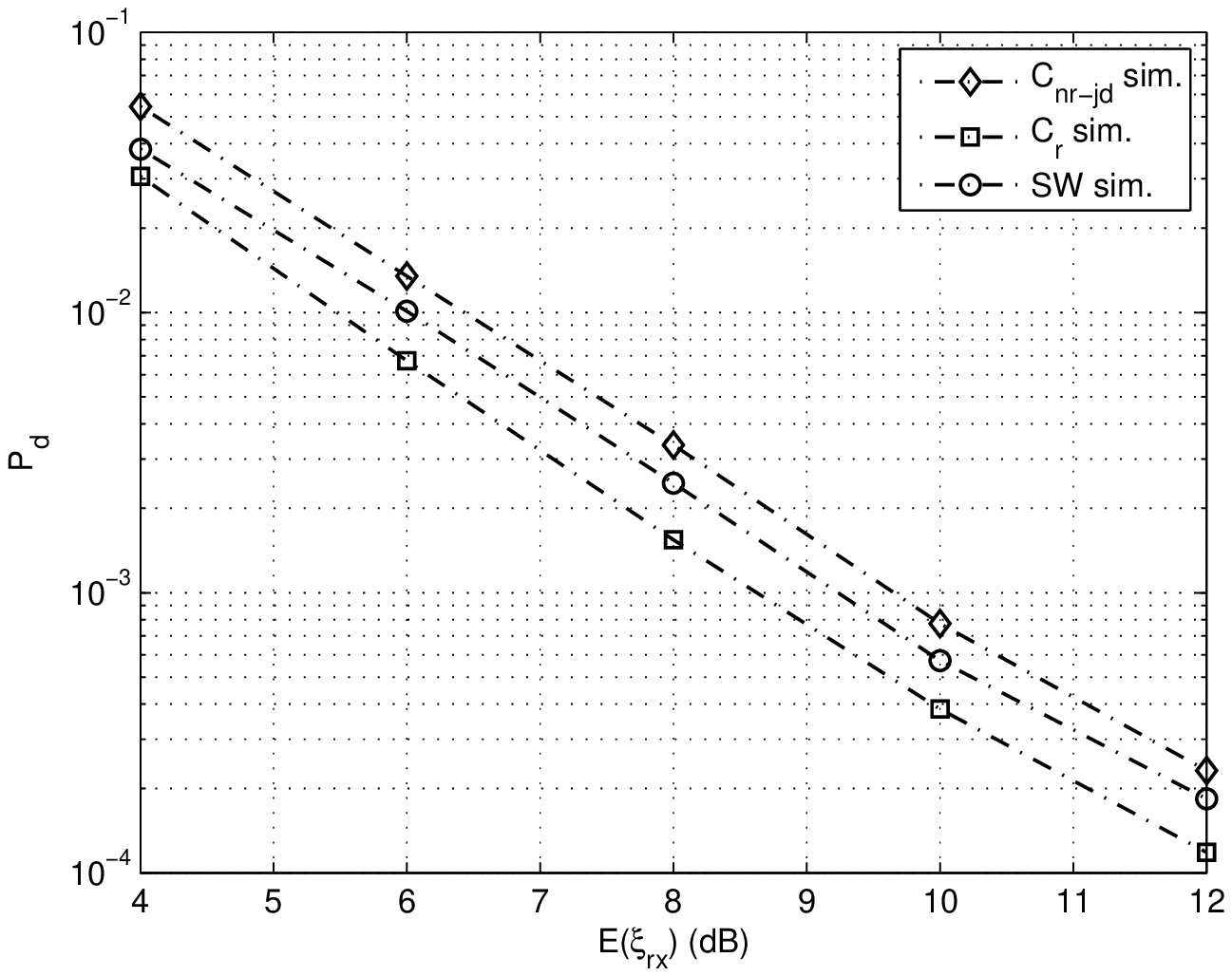}
\end{center} \caption{Comparison with the SW case: Rice channel model with $K_R = 10$} \label{Fig9}
\end{figure}

In order to test the effectiveness of the code searching strategy
shown in Section IV, computer simulations of the scenario proposed
in this paper have been carried out and comparisons with the
theoretical error bounds have been derived as well. In the simulated
scenario, channel decoding is based on the iterative approach
described in Section V. \\
The results are shown in Figs. \ref{Fig3}-\ref{Fig6}. In particular,
in Fig. \ref{Fig3} and \ref{Fig4} we set $\rho = 0.8$ while in Fig.
\ref{Fig5} and \ref{Fig6} we set $\rho = 0.9$. Besides, a packet
length $L_{pkt}=100$ is considered in Figs. \ref{Fig3} and
\ref{Fig5}, while a packet length $L_{pkt}=50$ is considered in
Figs. \ref{Fig4} and \ref{Fig6}. In the legend, sim. indicates
simulation results and bounds indicates theoretical bounds.
Different values of $\gamma_{b,x} = \gamma_{b,y}$ have been
considered in all Figs. and indicated in the abscissa as
$\gamma_{b}$. In the ordinate we have plotted the average packet
error probability ${P}_{d} = \frac{{P}_{d,x}+{P}_{d,y}}{2}$. In
these Figures we show results for the optimum recursive codes
reported in Table I, referred to as $C_{r}$, and for the $G^{(1)}(D)
= D^3+D^2+1$, $G^{(2)}(D) = D^3+D^2+D+1$ non-recursive code which is
optimum in the uncorrelated
 scenario \cite{Proakis}.
 Results obtained for the non-recursive code has been derived for both the joint detection and the unjoint detection case, and are referred to as $C_{nr-jd}$ and
 $C_{nr-ud}$, respectively \footnote{We do not use the same notation for the optimum recursive code $C_r$ since in this case we only perform joint
 detection. On the other hand, the unjoint detection case is equivalent to the uncorrelated case, where $C_{nr}$ is the optimum code.}. Unjoint detection means that the intrinsic correlation
 among information signals is not taken into account at the
 receivers and detection depicted in Figure \ref{Fig2i} is performed in only one
 step. In this case soft output measures are not necessary
 and, hence, we use a simple Viterbi decoder with hard output.\\
 Notice that, according to the analysis discussed in the previous Sections, the theoretical error bounds are expected to represent packet error probability's upper
 bounds (e.g., union bound probabilities). As a matter of fact,
 the theoretical bounds actually represent packet error probability's
 upper bounds for low packet error rates, when the
 assumption $\rho' = \rho$ is reasonable (\ref{eq3AAprev6bis}). Instead, for
 high packet error rates, i.e., for low $\gamma_b$, the theoretical
 bounds tend in some cases to superimpose the simulation curves. This is because
 for high bit error rates, i.e., for high packet error rates, the side-information is affected by non negligible errors and the
 hypothesis of perfect side information made in the analysis is not valid anymore. However, the theoretical bounds represent in all cases a good approximation
 of the simulation results. \\
 By observing again Figs. \ref{Fig3}-\ref{Fig6}, the following conclusions can be drawn. The optimum
 recursive codes allows to get an actual performance gain with respect to the non-recursive scheme, thus confirming the validity of the theoretical
 analysis described in previous Sections. Such a performance gain is
 particularly evident for high $\rho$ values, e.g., the performance gain at $P_d = 0.01$ is nearly of $0.6$ dB for $\rho = 0.9$ while for
 $\rho = 0.8$ the gain is less then $0.3$ dB. Comparisons with the unjoint detection case show
 that, as expected, joint detection allows to get a noticeable
 performance gain with respect to the unjoint case (from $0.6$ dB for
 $\rho = 0.8$ to more than $1.3$ dB for $\rho = 0.9$).\\
In order to assess the validity of the joint source-channel coding
approach considered in this paper, let's now provide a comparison
with a transmitting scheme which performs distributed source coding
achieving the Slepian-Wolf compression limit, and independent
convolutional channel coding. Note that such a scheme is ideal,
since the Slepian-Wolf compression limit cannot be achieved with
practical source coding schemes. For comparison purposes, we focus
on the $\rho = 0.9393$ case and we start by observing that the ideal
compression limit is equal to the joint entropy of the two
information signals $H({\mathbf{x}},{\mathbf{y}})$ =
$H({\mathbf{x}})$ + $H({\mathbf{x}}|{\mathbf{y}})$ = $1 - \rho
\times log_2(\rho) - (1-\rho)\times log_2(1-\rho)$ = $1.33$. In
order to get a fair comparison, let's now assume that the
transmitter with ideal Slepian Wolf compressor, referred to as $SW$
in the following, has at its disposal the same total energy and the
same transmitting time as the joint source-channel coding
transmitter without source compression proposed in this paper,
referred to as $JS-CC$ in the following. This means that the $SW$
transmitters can use the same energies $\xi_x$ and $\xi_y$ as the
$JS-CC$ transmitters and a reduced channel coding rates $r_{sw} =
\frac{1.33}{2}\times r = 2/3 r$, $r$ being the channel coding rate
for $JS-CC$. To be more specific, considering again $r=1/2$ for the
$JS-CC$ case, the $SW$ transmitting scheme can be modeled as two
independent transmitters which have to deliver $L_{pkt,sw}= 2/3
L_{pkt}$ independent information bits each one \footnote{Since the
$SW$ scheme performs ideal distributed compression, the original
correlation between information signals is fully lost}, using a
channel rate $r_{sw} = 1/3$ and transmitting energies $\xi_{x}$ and
$\xi_y$. As for the $JS-CC$ transmitting scheme, we consider both
the recursive $C_{95}$ channel coding scheme shown in Table I and
the $r = 1/2$ non-recursive coding scheme described above. As
before, the two cases are referred to as $C_r$ and $C_{nr-jd}$,
respectively. Note that in both cases we perform the iterative joint
decoding scheme described in the previous Section in an attempt to
exploit the correlation between information signals. Instead, since
distributed compression fully eliminates the correlation between
information signals, in the $SW$ case unjoint detection with hard
Viterbi decoding is performed at the receiver. As for the channel
coding scheme, we consider in the $SW$ case a non-recursive 1/3
convolutional code with $\nu = 3$ and with generator polynomials
$G^{(2)}(D) = D^3+D+1$, $G^{(2)}(D) = D^3+D^2+1$, $G^{(3)}(D) =
D^3+D^2+D+1$, \cite{Proakis}.\\
In order to provide an extensive set of comparisons between $C_r$,
$C_{nr-jd}$ and $SW$ we consider a more general channel model than
the AWGN considered so far. In particular, we assume that the link
gains $\alpha_x$ and $\alpha_y$ are RICE distributed \cite{Proakis}
with RICE factor $K_R$ equal to $0$ (i.e., Rayleigh case), $10$, and
$\infty$ (i.e., AWGN case). The three cases are shown in Figs.
\ref{Fig7}, \ref{Fig8} and \ref{Fig9}, respectively. We consider in
all cases a packet length $L_{pkt} = 100$. Moreover, we assume that
the two transmitters use the same transmitting energy per coded
sample $\xi = \xi_x = \xi_{y}$. In the abscissa we show the average
received power $E({\xi}_{rx}) = E\left(|\alpha_x|^2\right)\times
\xi_x = E\left(|\alpha_y|^2\right) \times \xi_{y}$ expressed in dB.
Note that the average $\gamma_b$ terms can be straightforwardly
derived as $E({\gamma}_b) = \frac{E({\xi}_{rx})}{2r} =
E({\xi}_{rx})$ for the $C_r$ and $C_{nr-jd}$ cases, and
$E({\gamma}_b) = \frac{E({\xi}_{rx})}{2 r_{sw}} = 1.5 \times
E({\xi}_{rx})$ for the $SW$ case. It is worth noting that the
comparisons shown in Figs. \ref{Fig7}, \ref{Fig8} and \ref{Fig9} are
fair in that $C_r$, $C_{nr-jd}$ and $SW$ use the same global energy
to transmit the same amount of information bits
in the same delivering time.\\
Notice from Fig. \ref{Fig7} that in the AWGN case $SW$ works better
than the other two schemes, even if the optimum recursive scheme
$C_r$ allows to reduce the gap from more then one dB to a fraction
of dB. The most interesting and, dare we say, surprising results are
shown in Figs. \ref{Fig8} and \ref{Fig9} where the $C_r$ decoding
scheme clearly outperform $SW$ with a gain of more then 1 dB in the
Rayleigh case and of almost 1 dB in the Rice case, while $C_{nr-jd}$
and $SW$ perform almost the same. This result confirms that, in
presence of many-to-one transmissions, separation between source and
channel coding is not optimum. The rationale for this result is
mainly because in presence of an unbalanced signal quality from the
two transmitters (e.g., independent fading), leaving a correlation
between the two information signals can be helpful since the better
quality received signal can be used as side information for
detecting the other signal. In other words, the proposed joint
decoding scheme allows to get a diversity gain which is not
obtainable by the $SW$ scheme. Such a diversity gain is due to the
inherent correlation between information signals and, hence, can be
exploited at the receiver without implementing any kind of
cooperation between the transmitters.

\section{Conclusions}

A simple wireless sensor networks scenario, where two nodes detect
correlated sources and deliver them to a central collector via a
wireless link, has been considered. In this scenario, a joint
source-channel coding scheme based on low-complexity convolutional
codes has been presented. Similarly to turbo or LDPC schemes, the
complexity at the decoder has been kept low thanks to the use of an
iterative joint decoding scheme, where the output of each decoder is
fed to the other decoder's input as a-priori information. For the
proposed convolutional coding/decoding scheme we have derived a
novel analytical framework for evaluating an upper bound of
joint-detection packet error probability and for deriving the
optimum coding scheme, i.e., the code which minimizes the packet
error probability. Comparisons with simulation results show that the
proposed analytical framework is effective. In particular, in the
AWGN case the optimum recursive coding scheme derived from the
analysis allows to clearly outperform classical non-recursive
schemes. As for the fading scenario, the proposed transmitting
scheme allows to get a diversity gain which is not obtainable by the
classical Slepian-Wolf approach to distributed source coding of
correlated sources. Such a diversity gain allows the proposed scheme
to clearly outperform a Slepian-Wolf scheme based on ideal
compression of distributed sources.


\newpage
\bibliographystyle{IEEE}

\end{document}